\documentclass[letterpaper, 10 pt, conference]{ieeeconf}
\usepackage{url}
\usepackage{booktabs}
\usepackage{calc}
\usepackage{graphicx}
\usepackage{multirow}
\usepackage[ruled, vlined, longend, linesnumbered]{algorithm2e}
\DontPrintSemicolon
\SetKwInOut{Input}{Input}
\SetKwInOut{Output}{Output}
\SetAlFnt{\small}
\newlength{\commentWidth}
\SetCommentSty{mycommfont}

\usepackage{comment}
\usepackage{paralist}
\usepackage{color}
\usepackage{paralist}
\usepackage{xcolor}
\usepackage{amsmath}
\usepackage{amssymb}

\usepackage{amsthm}

\usepackage{bm}
\usepackage{xspace}
\let\labelindent\relax
\usepackage{enumitem}
\usepackage{pifont}
\usepackage[labelformat=simple, font=small, skip=3pt]{subcaption}
\usepackage[font=small, skip=3pt]{caption}
\usepackage[utf8]{inputenc}
\usepackage{csquotes}
\usepackage{graphicx}
\usepackage{soul,color}
\usepackage{lipsum}
\usepackage[noadjust]{cite}
\usepackage[subpreambles=true]{standalone}

\graphicspath{{figures/}}

\makeatletter
\newcommand*{\rom}[1]{\expandafter\@slowromancap\romannumeral #1@}
\makeatother

\newtheorem{theorem}{Theorem}

\theoremstyle{definition}

\newtheorem{problem}{Problem}

\setlist[description]{font=\normalfont\itshape\textbullet\space}

\newcommand{\R}{\mathbb{R}}

\DeclareMathOperator*{\argmax}{arg\,max}

\newcommand{\vcycle}{\ensuremath{V\textsubscript{cycle}}\xspace}
\newcommand{\val}{\ensuremath{val}\xspace}
\newcommand{\opt}{\ensuremath{opt}\xspace}

\newcommand{\ofr}{\textsc{OFr}\xspace}
\newcommand{\ofri}{\textsc{OFr-I}\xspace}
\newcommand{\osc}{\textsc{OSc}\xspace}
\newcommand{\hgc}{\textsc{HGc}\xspace}

\newcommand{\gfr}{\textsc{GFr}\xspace}

\newcommand{\gpr}{\textsc{GPr}\xspace}

\newcommand{\prob}{COP\xspace}

\begin{document}
\title{\LARGE \bf Optimal Routing Schedules for Robots Operating in Aisle-Structures}
\author{
	Francesco Betti Sorbelli,
	Stefano Carpin,
	Federico Cor\`o,
	Alfredo Navarra,
	and Cristina M.~Pinotti
	\thanks{
		Francesco Betti Sorbelli, Alfredo Navarra, and Cristina M.~Pinotti are with the 
		University of Perugia, Italy. 
		Stefano Carpin is with the 
		University of California, Merced, CA, USA.
		Federico Cor\`o is with the 
		Gran Sasso Science Institute (GSSI), L'Aquila, Italy.
	}
}

\maketitle

\bstctlcite{IEEEexample:BSTcontrol}

\begin{abstract}
In this paper, we consider the Constant-cost Orienteering Problem (\prob)
where a robot, constrained by a limited travel budget, aims at selecting a path with the largest reward in an aisle-graph.
The aisle-graph consists of a set of  loosely connected rows where
the robot can change lane only at 
either end, but not in the middle.
Even when considering this special type of graphs, the orienteering problem is known to be $\mathit{NP}$-hard.
We optimally solve in polynomial time two special
cases, \prob-FR where the robot can only traverse full rows,
and \prob-SC where the robot can access the rows only from one side.
To solve the general \prob, we then apply our special case algorithms 
as well as a new heuristic that suitably combines them.
Despite its light computational complexity and being confined into a very limited class of paths,
the optimal solutions for \prob-FR turn out to be competitive even for \prob in both real and synthetic scenarios.
Furthermore, our new heuristic for the general case outperforms  state-of-art algorithms,
especially for input with highly unbalanced rewards. 
\end{abstract}

\section{Introduction}
\begin{figure}[h] 
	\centering
	\begin{subfigure}[b]{0.5\textwidth}
		\centering
		\includegraphics[width=8.00cm]{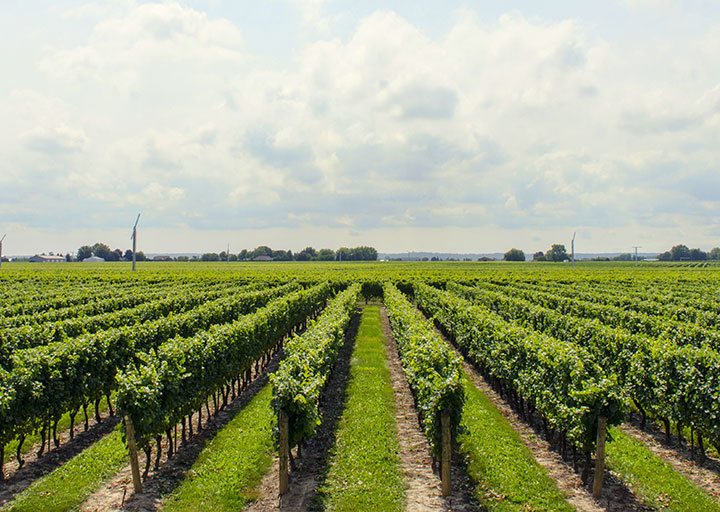}
		\caption{Vineyard.}
		\label{img:vineyard}
	\end{subfigure}
	\begin{subfigure}[b]{0.5\textwidth}
		\centering
		\includegraphics[width=8.00cm]{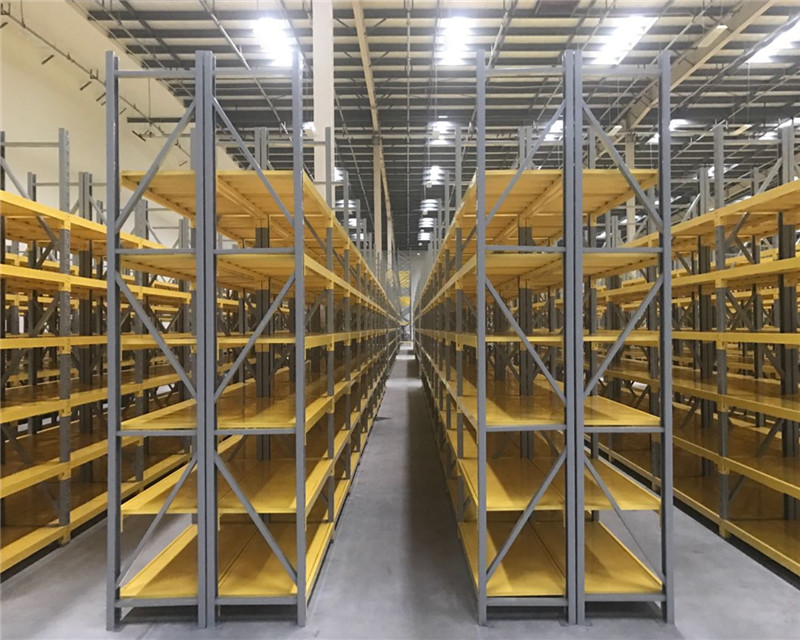}
		\caption{Warehouse.}
		\label{img:shelves}
	\end{subfigure}
	\caption{Real examples of aisle-structures.}
	\label{img:applications}
\end{figure}

Numerous robotic tasks can be abstracted using a graph model featuring 
vertices associated with rewards and edges associated with costs. 
Vertices usually represent locations that the robots must visit to perform some task,
while edges costs represent the energy or time spent while moving between locations. 
Models like these emerge for example when robots are used for
environmental monitoring, surveillance, logistic, urban mobility, and
precision agriculture, just to name a few.
Owing to the fact that most robots and vehicles have to periodically 
stop to recharge their batteries or refuel, this type of tasks are naturally
connected to the combinatorial optimization problem known in literature as \emph{orienteering}. 
In orienteering one is given a graph where each vertex has a reward and each edge has a cost.
The objective is to determine a path on the graph maximizing the sum of collected
rewards while ensuring that the cost of the path does not exceed a given budget. 
An important feature of this problem is that if a vertex
is visited more than once the reward is collected only once, while the cost
associated with an edge is incurred every time it is traversed. 
As pointed out in~\cite{thayer2018routing}, this optimization problem
is computationally hard, and therefore it is of interest to study either
heuristic solutions that perform well on large problem instances, or
exact solutions when one considers special types of graphs or specific 
classes of paths. In~\cite{thayer2018routing} the orienteering problem was used
to solve  a routing problem associated with robots used for
precision irrigation in vineyards.
A vineyard imposes specific 
motion constraints because a robot operating in it can change
row only when it is at either end of the row, but not when it is in the middle.
The same motion constraints can be found in warehouses,
which are typically organized with
long rows of parallel shelves, and crossing in the middle
is typically impossible due to stored goods~\cite{bettisorbelli2019automated}. 
Fig.~\ref{img:applications} illustrates the two scenarios above.
These motion constraints can be captured by the \emph{aisle-graph} (a.k.a. \emph{irrigation graph}~\cite{thayer2018routing}),
formally defined in Sec.~\ref{sec:definition}.
On such graphs, an associated orienteering
problem can be defined to account for the limited battery runtime of the robot.
In~\cite{thayer2018routing} it was shown that solving the orienteering problem
on the special class of aisle-graphs is $\mathit{NP}$-hard, and two different heuristics
were proposed. 
In this work we propose various improvements to the results 
presented in~\cite{thayer2018routing}. First, we provide an efficient, optimal 
algorithm to solve the orienteering problem on graphs when one restricts 
the solution to a specific class of paths called \emph{full rows}, that were
heuristically explored in~\cite{thayer2018routing}. Then, we introduce
a new class of paths called \emph{single column} that is related to one of the
heuristics given in~\cite{thayer2018routing}, and for this case we also provide
an efficient, optimal solution. Finally, building upon the two optimal solutions found,
we introduce a new heuristic that performs well in practical different scenarios.
To evaluate the strength of our findings, we test our algorithms
on synthetic instances based on various Zipf distributions~\cite{tullo2003modelling}, 
and on the same benchmarks proposed in~\cite{thayer2018routing} that are generated from
a real world robotic precision irrigation application with graphs featuring 
more than 50,000 vertices.

\subsection{Related Work}\label{sec:related}
The \emph{orienteering} problem was first formalized and
studied in~\cite{OrienteeringProblemGolden}, where its $\mathit{NP}$-hardness was also
proven. Later, Blum et al. showed that the problem is $\mathit{APX}$-hard~\cite{BlumOrienteering}.
Due to its inherent computational complexity, numerous heuristic approaches are found
in literature, and due to space constraints, the reader is referred to~\cite{orienteeringSurvey2016} for a
comprehensive overview of heuristics and problem variants.
A distinct line
of research aims instead at solving the orienteering using exact formulations~\cite{FischettiOrienteering}.
These solutions, however, do not scale to instances as large as those we consider
in this work.

Besides its theoretical interest, the orienteering problem finds
applications also in robotics and automation~\cite{JorgensenChenEtAl2017b,thayer2018arouting,RusTRO2016,SaskaRAL2019}.
In~\cite{thayer2018routing}, it was shown that robotic routing problem described above in aisle-graphs
can be formulated in terms of orienteering, and the most recent
greedy solutions ad-hoc for aisle-graphs were proposed.
In~\cite{bettisorbelli2019automated}, the aisle-graphs are used to model a warehouse where 
an automated picking system is implemented. The warehouse has cabinets that form long lanes separated by aisles.
In the automated picking system,
the robot has to collect all the items requested by a customer order
minimizing the distance (i.e., budget) traversed and changing lane only at the lane extreme, not in the middle. 
The picking problem has been solved by applying the well-known Christofides' algorithm 
for Traveling Salesman Problem~\cite{christofides1976worst}.

\paragraph*{Organization}
The remainder of this paper is organized as follows.
The formal problem definition is provided in Sec.~\ref{sec:definition}. 
Our novel algorithms are presented in Sec.~\ref{sec:solutions}, together with their complexity analysis.
In Sec.~\ref{sec:simulations} we evaluate the effectiveness of our approach on large scale instances, and
in Sec.~\ref{sec:conclusions} we draw conclusions and sketch avenues for future work.

\section{Problem Definition}
\label{sec:definition}
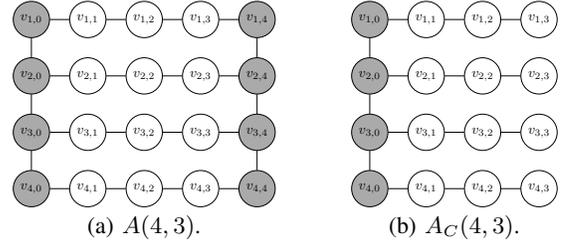
\begin{figure}[htbp]
	\centering
	\begin{subfigure}[b]{0.26\textwidth}
		\centering
		\begin{tikzpicture}[scale=1, every node/.style={scale=0.7}]
    
    \node[circle, draw, text=black, minimum size=0.5cm, fill={rgb:black,1;white,2}] (A)   at (0,0)  {$v_{4,0}$};
    \node[circle, draw, text=black, minimum size=0.5cm] (B)   at (1,0) {$v_{4,1}$};
    \node[circle, draw, text=black, minimum size=0.5cm] (C)   at (2,0) {$v_{4,2}$};
    \node[circle, draw, text=black, minimum size=0.5cm] (D)   at (3,0) {$v_{4,3}$};
    \node[circle, draw, text=black, minimum size=0.5cm, fill={rgb:black,1;white,2}] (E)   at (4,0)  {$v_{4,4}$};
    
    \node[circle, draw, text=black, minimum size=0.5cm, fill={rgb:black,1;white,2}] (F)   at (0,1)  {$v_{3,0}$};
    \node[circle, draw, text=black, minimum size=0.5cm] (G)   at (1,1)  {$v_{3,1}$};
    \node[circle, draw, text=black, minimum size=0.5cm] (H)   at (2,1)  {$v_{3,2}$};
    \node[circle, draw, text=black, minimum size=0.5cm] (I)   at (3,1)  {$v_{3,3}$};
    \node[circle, draw, text=black, minimum size=0.5cm, fill={rgb:black,1;white,2}] (L)   at (4,1)  {$v_{3,4}$};
    
    \node[circle, draw, text=black, minimum size=0.5cm, fill={rgb:black,1;white,2}] (M)   at (0,2)  {$v_{2,0}$};
    \node[circle, draw, text=black, minimum size=0.5cm] (N)   at (1,2)  {$v_{2,1}$};
    \node[circle, draw, text=black, minimum size=0.5cm] (O)   at (2,2)  {$v_{2,2}$};
    \node[circle, draw, text=black, minimum size=0.5cm] (P)   at (3,2)  {$v_{2,3}$};
    \node[circle, draw, text=black, minimum size=0.5cm, fill={rgb:black,1;white,2}] (Q)   at (4,2)  {$v_{2,4}$};
    
    \node[circle, draw, text=black, minimum size=0.5cm, fill={rgb:black,1;white,2}] (R)   at (0,3)  {$v_{1,0}$};
    \node[circle, draw, text=black, minimum size=0.5cm] (S)   at (1,3)  {$v_{1,1}$};
    \node[circle, draw, text=black, minimum size=0.5cm] (T)   at (2,3)  {$v_{1,2}$};
    \node[circle, draw, text=black, minimum size=0.5cm] (U)   at (3,3)  {$v_{1,3}$};
    \node[circle, draw, text=black, minimum size=0.5cm, fill={rgb:black,1;white,2}] (V)   at (4,3)  {$v_{1,4}$};
        
    \draw (A) -- (B) -- (C) -- (D) -- (E);
    \draw (F) -- (G) -- (H) -- (I) -- (L);
    \draw (M) -- (N) -- (O) -- (P) -- (Q);
    \draw (R) -- (S) -- (T) -- (U) -- (V);
    
    \draw (A) -- (F) -- (M) -- (R);
    \draw (E) -- (L) -- (Q) -- (V);
    
\end{tikzpicture}
		\caption{$A(4, 3)$.}
		\label{fig:example-IG}
	\end{subfigure}
	\begin{subfigure}[b]{0.19\textwidth}
		\centering
		\begin{tikzpicture}[scale=1, every node/.style={scale=0.7}]

\node[circle, draw, text=black, minimum size=0.5cm, fill={rgb:black,1;white,2}] (A)   at (0,0)  {$v_{4,0}$};
\node[circle, draw, text=black, minimum size=0.5cm] (B)   at (1,0) {$v_{4,1}$};
\node[circle, draw, text=black, minimum size=0.5cm] (C)   at (2,0) {$v_{4,2}$};
\node[circle, draw, text=black, minimum size=0.5cm] (D)   at (3,0) {$v_{4,3}$};

\node[circle, draw, text=black, minimum size=0.5cm, fill={rgb:black,1;white,2}] (F)   at (0,1)  {$v_{3,0}$};
\node[circle, draw, text=black, minimum size=0.5cm] (G)   at (1,1)  {$v_{3,1}$};
\node[circle, draw, text=black, minimum size=0.5cm] (H)   at (2,1)  {$v_{3,2}$};
\node[circle, draw, text=black, minimum size=0.5cm] (I)   at (3,1)  {$v_{3,3}$};

\node[circle, draw, text=black, minimum size=0.5cm, fill={rgb:black,1;white,2}] (M)   at (0,2)  {$v_{2,0}$};
\node[circle, draw, text=black, minimum size=0.5cm] (N)   at (1,2)  {$v_{2,1}$};
\node[circle, draw, text=black, minimum size=0.5cm] (O)   at (2,2)  {$v_{2,2}$};
\node[circle, draw, text=black, minimum size=0.5cm] (P)   at (3,2)  {$v_{2,3}$};

\node[circle, draw, text=black, minimum size=0.5cm, fill={rgb:black,1;white,2}] (R)   at (0,3)  {$v_{1,0}$};
\node[circle, draw, text=black, minimum size=0.5cm] (S)   at (1,3)  {$v_{1,1}$};
\node[circle, draw, text=black, minimum size=0.5cm] (T)   at (2,3)  {$v_{1,2}$};
\node[circle, draw, text=black, minimum size=0.5cm] (U)   at (3,3)  {$v_{1,3}$};
        
    \draw (A) -- (B) -- (C) -- (D); 
    \draw (F) -- (G) -- (H) -- (I); 
    \draw (M) -- (N) -- (O) -- (P); 
    \draw (R) -- (S) -- (T) -- (U); 
    
    \draw (A) -- (F) -- (M) -- (R);
    
\end{tikzpicture}
		\caption{$A_C(4, 3)$.}
		\label{fig:example-IG-one-column}
	\end{subfigure}
	\caption{Examples of graphs: the interconnecting vertices are dark.}
	\label{fig:examples-IG}
\end{figure}

Let us consider an undirected {\em Aisle Graph} $A(m, n) = (V, E)$, 
where $m$ and $n$ denote the number of rows and columns, respectively.
We define the set of vertices $V = \{ v_{i,j} | 1 \le i \le m, 1 \le j \le n \} \cup v_{i,0},v_{i,n+1} \forall i\in 1, \ldots, n$ 
and the set of edges $E$ is built as follows:
\begin{itemize}
	\item Each vertex $v_{i,j}$ with $1 \le i \le m$ and $1 \le j \le n$ has two edges, 
	one toward $v_{i,j-1}$ and the other toward $v_{i,j+1}$;
	\item each vertex $v_{i,0}$ with $1 < i < m$ has three edges: one toward $v_{i-1,0}$, 
	one toward $v_{i+1, 0}$, and one toward $v_{i,1}$;
	symmetrically,  each vertex of type $v_{i,n+1}$ with $1 < i < m$ has three edges: 
	one toward $v_{i-1,n+1}$, one toward $v_{i+1, n+1}$, and one toward $v_{i,n}$.
	Accordingly, edges connected to the corner vertices $v_{1,0}$, $v_{m,0}$, $v_{1,n+1}$ and $v_{m,n+1}$ are then well defined.
\end{itemize}
The graph $A(m, n)$  has therefore $m$ rows and $n+2$ columns.
Note that, the vertices in the first and last columns (e.g., Fig.~\ref{fig:example-IG} with indices $j=0$ and $j=n+1$, respectively) do not provide any reward, 
and their purpose is just to connect the $m$ rows.

\begin{problem}[Constant-cost Orienteering Problem (\prob)]
	Let $A(m, n)$ be an aisle-graph, $v_s, v_d \in V$ be two of its vertices, $r : V \rightarrow \R_{\ge 0}$ be a reward function where $r(v_{i,j})=0$ if $j\in\{0,n+1\}$, 
	and $c : E \rightarrow \R_{\ge 0}$ be a constant cost function on $A$, i.e., $c(e) = \alpha$ for each $e \in E$. 
	The \prob asks, for a given constant $B$, to find a path of maximum reward starting at $v_s$ and ending at $v_d$ of cost no greater than $B$.
\end{problem}

W.l.o.g., it can be considered the case $c(e) = 1$ for each $e \in E$, whereas for our purposes we restrict to the case $v_s = v_d = v_{1, 0}$.
In what follows, by $r_i$ and $c_j$ we denote the $i$-th row and the $j$-th column of $A$, respectively,
and by $ R = \{ r_1, \ldots, r_{m} \}$ and $ C = \{ c_0, \ldots, c_{n+1} \}$ the set of rows and columns of $A$, respectively.

Moreover, we define an undirected {\em single column} graph  $A_C(m, n) = (V, E)$ as an Aisle Graph which interconnects the $m$ rows just using $c_0$. 
The set of vertices $V$ and the set of edges $E$ are equal to those of $A(m,n)$ with the only exception of column $c_{n+1}$ which is missing in $A_C$ and thus the vertices of type $v_{i,n}$, with $1 \le i \le m$,
admit just one edge toward $v_{i,n-1}$.
An instance of $A_C$ is shown in Fig.~\ref{fig:example-IG-one-column}.

\section{Proposed Algorithms}
\label{sec:solutions}
In this section, we first optimally solve two special cases of \prob:
\prob-FR (full row)  which forces the robot to entirely traverse rows in $A$,
and \prob-SC (single column) which optimally solves \prob on $A_C$.

\prob-FR is a direct abstraction of real-world scenarios in which the robot can neither turn around nor
reverse its movement.
For example, if we are representing a warehouse with very narrow aisles or environments with multiple robots working simultaneously, in that case reversing the direction may be prohibited to avoid crashes, or impossible due to maneuvering limits. 
On the other hand, \prob-SC may model instead robot motions in combed-shaped or star-shaped structures.
Moreover, \prob-SC can be easily modified to be used to solve \prob with two robots working in parallel: one on the left side and one on the right side of an aisle-graph.

Lastly, we provide the \hgc heuristic for solving \prob on an aisle-graph by suitably combining the aforesaid optimum strategies.

\subsection{Optimal Solutions for \prob-FR}\label{sec:full-rows-policy}
In this section we propose two polynomial algorithms, called \ofr and \ofri, to solve \prob-FR.
The latter is an efficient  implementation of \ofr.
Both algorithms return a subset $S$ of rows as a solution, i.e., $ S \subseteq R $, such that the path starting and ending at $v_{1, 0}$ has cost no greater than $B$ and collects  maximum reward.

The following properties hold for \prob-FR:
\begin{description}
	\item[P1:] A reasonable budget $B\ge 2(n+1)$ must hold
    since a smaller value makes impossible to traverse any row of the graph and coming back to $v_{1, 0}$.
    In contrast if $B \ge (n+1)m+2(m-1)$ then the problem becomes trivial since 
    the robot can easily perform a complete visit of the graph.
    \item[P2:] In any optimal solution $S$, the robot performs a cycle traversing the rows in $S$ in an increasing (or decreasing) order with respect to their indices.
    Any other cycle would require a larger budget.
    \item[P3:] Any optimal solution $S$ that starts and ends at $v_{1, 0}$ must fully traverse an even number of rows to prevent the robot to get stuck in $c_{n+1}$.
\end{description}

\subsubsection{The \ofr Algorithm}
We now introduce the \textsc{Opt Full-row} (\ofr) algorithm that optimally solves \prob-FR.
The pseudo-code of \ofr is provided in Algorithm~\ref{alg:opt-full-row}.

\begin{algorithm}[ht]
    $\vcycle \gets \emptyset, \opt \gets -\infty$\;
    $R_i \gets \sum_{j=1}^{n} r(v_{i, j}) \;\forall i \in 1, \ldots, m$\label{code:cum-rewards}
    $, \tilde{R} \gets \textsc{sort}(R_i)$\label{code:sort}\;
	\For{$ m' \gets 1, m $}{\label{code:farthest-row}
	    $V = \emptyset, \val \gets 0$\;
	    $B' \gets B - 2(m'-1)$\;\label{code:residual-budget}
	    \If(\tcp*[h]{Build current solution}){$B'>0$}{
	        $k(m')=2 \left \lfloor \frac{B'}{2(n+1)} \right \rfloor$ \;\label{code:permitted-rows}
            \If{$k(m') \ge m'$}{\label{code:budget-ok}
                $\val \gets \sum_{j=1}^{m'}R_j, \textsc{updateTour}(V)$\;
	        }\Else{
		        $j \gets 1, s \gets 1$\;
		        $\val \gets  R_{m'}$\;
                \While{$s \le k(m')-1$}{\label{code:test-rows}
                    \If{$i_{j} < m'$}{
                        $\val \gets  \val+\tilde{R}_{i_{j}}, \textsc{updateTour}(V)$\;
                        $s \gets  s+1$\;
                    }
                    $j \gets  s+1$\;
		        }
		    }
		    \If{$\val > \opt$}{\label{code:update-opt}
		        $\vcycle \gets V, \opt \gets \val$\;
		    }
		}
	}
\caption{\textsc{\ofr}\(( \text{Graph } A, \text{Budget } B)\)}
\label{alg:opt-full-row}
\end{algorithm}

\ofr consists of two steps. In the first pre-processing step,
for each row $i$
a cumulative reward $R_i$, i.e., the reward collected by completely traversing the $i$-th row, is computed (Line~\ref{code:cum-rewards}), 
as $R_i = \sum_{j=1}^{n} r(v_{i, j})$.
Then (Line~\ref{code:sort}), we sort  the $R_i$ values in  decreasing order obtaining the row 
permutation $\tilde{R}= [{i}_1, \ldots, {i}_m]$ such that
$\tilde{R}_{i_1} \ge \ldots \ge \tilde{R}_{i_m}$.
This phase has cost $\mathcal{O}(mn + m \log m)$.

In the second step, exploiting the three properties above,
the optimal solution with budget $B$ can be easily determined if we fix the
furthest row traversed by the algorithm.

Letting $m'$ be the furthest traversed row (Line~\ref{code:farthest-row}), 
the solution moving vertically spends $(m'-1)$ units of budget
to reach row $m'$ from $v_{1,0}$, 
and additional $(m'-1)$ units to go back to $v_{1,0}$.
Any optimal solution can either first visit the selected rows up to row $m'$
and then come back, or
first reach the furthest row $m'$
and then come back visiting the selected rows.
The two options have the same vertical cost of $2(m'-1)$:
the vertical movements of the robot from row $1$ to row $m'$  alternate on $c_0$ and $c_{n+1}$. 
In the former case, the vertical movement from row $1$ to the first row selected in the solution
are performed on $c_0$.
Once the first row is reached, the robot traverses it left to right,  ends up in $c_{n+1}$,
and moves vertically from the first
to the second selected row on $c_{n+1}$.
Since the second row is traversed right to left, the robot ends up in $c_0$,
and starts again going down to reach the third selected row.
After traversing the furthest row $m'$, if $m'$ is even, the robot goes back to $v_{1,0}$
traversing $c_0$. Otherwise, if $m'$ is odd, the robot 
traverses row $m'$ again from right to left
and reaches $v_{1,0}$ moving on $c_0$.

Given that $2(m'-1)$ is the budget required for the vertical movements,
the residual budget $B' = B-2(m'-1)$ (Line~\ref{code:residual-budget})
is  spent to traverse as many even rows as possible (Line~\ref{code:permitted-rows})
that is
$k(m')=2 \left \lfloor \frac{B'}{2(n+1)} \right \rfloor$.
The optimal solution than includes row $m'$ and the $k(m')-1$ rows with the largest reward
among the first $m'$ rows
(Line~\ref{code:test-rows}) if $k(m') \le m'$.
If $k(m') > m'$ (Line~\ref{code:budget-ok}), then all the $m'$ rows are selected. If $m'$ is odd, to make an even number of rows, a row, say $m'$ for the sake of simplicity, is traversed twice.
Note that, if $B < 2(m'-1)$, the $m'$-th row cannot be reached
and there is no feasible solution that includes row $m'$.

To find the optimal solution constrained to budget $B$,
we compute the most profitable solution for any possible value of $m'$ (Line~\ref{code:farthest-row})
and we retain the solution with the largest overall reward (Line~\ref{code:update-opt}).

Finally, the procedure $\textsc{updateTour}(V)$ constructs the path accordingly to the rows
that the robot has to take, and adding the suitable vertical movements as explained above.

The total computational cost of the algorithm is $\mathcal{O}(m\cdot \max\{n,m\})$ since
it spends $\mathcal{O}(mn + m \log m)$ for the pre-processing phase
and $\mathcal{O}(m^2)$ for the second phase.

By the above discussion the next theorem can be stated. 

\begin{theorem}\label{alg:algoOFR}
Algorithm \ofr
optimally solves \prob-FR in time $\mathcal{O}(m\cdot \max\{n,m\})$.
\end{theorem}

 \begin{proof}
 About the complexity of the algorithm, it has been already analyzed above.

 Given a graph $A(m,n)$ and a budget $B$, 
 consider the optimal solution given as the set of rows traversed by the robot, $OPT = \{ r_{i_1}, \ldots, r_{i_{k^*}} \}$ of size $k^*$. By property P3, $k^*$ is even.  
 By P2,  $r_{i_j} \le r_{i_{j+1}}$ for any $j<k^*$, i.e., the rows are increasingly ordered with respect to their index.
 Now consider Algorithm~\ref{alg:opt-full-row} at iteration $m' = i_{k^*}$.
 To reach the $m'$-th row and go back, at least $2(m-1)$ units of budget have to be spent.
 Let $k=k(m')$ be the number of rows defined at Line~\ref{code:permitted-rows} that will be visited by the robot
 with the residual budget $B'$. 
 Since $k(m')=2 \left \lfloor \frac{B'}{2(n+1)} \right \rfloor$ is even and
 since
 there is not enough budget to traverse any other pair of rows given that  $B-k(m')(n+1)<2(n+1)$,
 $k(m')$ is the largest possible even number of full rows when the solution includes  row $m'$.
  If $k(m') > m'$ and $m'$ is odd,
  then the sub-procedure $\textsc{updateTour}(V)$ will make the robot traverse row $m'$ twice.
 Finally, fixed $m'$ and computed $k(m')$, the solution is optimal because it selects the most profitable rows
 (see Line~\ref{code:test-rows}) 
 according to
 the sorting of Line~\ref{code:sort}.
 Since all the possible values of $m'$ are tested, Algorithm~\ref{alg:opt-full-row} optimally solve \prob-FR.
\end{proof}

\subsubsection{The \ofri Algorithm}
We next present a faster implementation for \ofr, called \ofri.
The idea is to reduce the cost of selecting the rows of the optimal solution from $\mathcal{O}(m^2)$ time for \ofr to $\mathcal{O}(m)$ time for \ofri.
The pseudo-code of \ofri is provided in Algorithm~\ref{alg:opt-full-row-improved}.

\begin{algorithm}[htbp]
	\caption{\textsc{\ofri}\(( \text{Graph } A, \text{Budget } B)\)}
	\label{alg:opt-full-row-improved}
	$\vcycle \gets \emptyset,\; \opt \gets -\infty,\; old \gets 0$\;
	$S \gets [1 \ldots m]$, $j \gets 1 \quad$\tcp{Current index}
    $R_i \gets \sum_{j=1}^{n} r(v_{i, j}) \;\; \forall i \in 1, \ldots, m, \tilde{R} \gets \textsc{sort}(R)$\;
	\For(\tcp*[h]{Decreasing cycle}){$ m' \gets m, 1 $}{\label{code:2-first-row}
		$V = \emptyset, \val \gets 0$\;
	    $B' \gets B - 2(m'-1)$\;
	    $k(m')=2 \left \lfloor \frac{B'}{2(n+1)} \right \rfloor$ \;\label{code:2-permitted-rows}
	    \If(\tcp*[h]{$1$-st iteration}){$m' = m$}{\label{code:2-first-iteration}
	    	\While{$ j \le k(m')$}{\label{code:2-first-km-rows}
	    		$S[i_j] \gets 1$\;\label{code:2-first-add}
	    		$\val \gets  \val+\tilde{R}_{i_{j}}, \textsc{updateTour}(V)$\;
        		$j \gets j+1\qquad$\tcp{$j$ scans $\tilde{R}$}
        	}
			$old \gets k(m')$\;\label{code:2-first-old}
		}\ElseIf{$m' < k(m')$}{\label{code:2-last}
			\For{$i \gets 1, m'$}{
                $S[i] \gets 1$\;
                $\val \gets  \val+{R}_{i}, \textsc{updateTour}(V)$\;
            }
            $m' \gets 1$\;
        }\Else{\label{code:2-general-condition}
        	$c \gets 0$\;
        	\If{$k(m') \neq old$}{\label{code:2-general-two-more}
			    $c \gets 2$\;
		    }
		    \If{$S[m'+1]=1$}{\label{code:2-general-delete}
			    $c \gets c+1, S[m'+1] \gets 0$\;
			    $\val \gets  \val - {R}_{m'+1}, \textsc{updateTour}(V)$\;
		    }
		    $t \gets 1$\;
		    \While(\tcp*[h]{Finds $c$ feasible rows}){$t \leq c$}{\label{code:2-general-add}
				\If{$i_j \le m'$}{
					$S[i_j] \gets 1$\;
					$\val \gets  \val+\tilde{R}_{i_{j}}, \textsc{updateTour}(V)$\;
			        $t \gets t+1$\;
				}
				$j \gets j+1$\;
			}
			$old \gets k(m')$\;\label{code:2-general-old-updated}
        }
        \If{$\val > \opt$}{\label{code:2-update-opt}\label{code:2-general-current-solution-vs-opt}
		    $\vcycle \gets V, \opt \gets \val$\;
		}
    }
\end{algorithm}

Initially, \ofri performs the same pre-processing phase as the one of \ofr.
Then, in the second step, 
the optimal solution is calculated in the opposite way with respect to \ofr.
In fact, we first set $m'$ as the farthest possible row, i.e., $m'=m$ (Line~\ref{code:2-first-row}).
Then, the main loop proceeds decreasing $m'$ and, at each step, according to the value of $m'$,
the residual budget and the number of permitted rows to be considered are
computed (Line~\ref{code:2-permitted-rows}).
During the first iteration of the main loop, i.e., $m'=m$ (Line~\ref{code:2-first-iteration}), 
the algorithm builds the current solution $S$ by picking the first $k(m')$ rows in $\tilde{R}$
(Line~\ref{code:2-first-km-rows}).
The indices of the top $k(m)$ rows with the largest rewards (Line~\ref{code:2-first-km-rows})
are inserted in the Boolean vector $S$ (Line~\ref{code:2-first-add}).
Precisely, $S[i_j]=1$ if $i_j$ is one of the first $k(m)$ entries in $\tilde{R}$, 
and $S[i_j]=0$ otherwise,
with $1 \le i_j \le m$.
Clearly, the insertion procedure can be done in constant time for each row.
Note that using this approach we are not forcing row $m'$ to belong to the current solution.
It is worthy to note that when $m'=m$  the total reward obtained by \ofri  is larger or equal
to that of \ofr constrained to use row $m$ because \ofri takes the best possible top $k(m)$ rows and \ofr only the 
top $k(m)-1$ best possible rows plus the $m$-th.
When $m' < m$,
\ofri evaluates the new residual budget.
As in \ofr, the best solution using the rows from $1$ to $m'$ is searched.
It may happen that 
\begin{inparaenum}
    \item the number of permitted rows $k(m')$ increases with respect to $k(m'+1)$, and
    \item the row $m'+1$ has to be removed from the current solution to be substituted by a row with index no larger than $m'$. 
\end{inparaenum}
In the first case, 
the number of permitted rows may increase
since  the
length of the vertical movements decreases by $2$ and thus, the residual budget increases by $2$.
According to Line~\ref{code:2-permitted-rows}, it must hold $c=k(m'+1)-k(m')=2$.
In the second case the algorithm is able to remove the $(m'+1)$-th row in constant time just by setting $S[m'+1]=0$ and increasing by $1$ the number $c$ of rows to be added to the current solution (Line~\ref{code:2-general-delete}).

For every value of $m'$,
the algorithm adds $c$ rows scanning $\tilde{R}$ starting from the position $j$
of the last row inserted and 
ignoring the rows with index larger than $m'$ (Line~\ref{code:2-general-add}).
If $c=0$, the solution $S$ is unchanged
and the algorithm proceeds decreasing $m'$.
Otherwise, $\tilde{R}$ is scanned from position $j$ to find $c$ new rows.
Note that $j$ never decreases because any row discarded during a previous iteration cannot be inserted in this iteration either.
Namely, whenever a row $i_j$ in $\tilde{R}$ is discarded from $S$ it is because
its row index is greater than the current value $m'$, i.e., $i_j > m'$,
and since $m'$ decreases, $i_j$ cannot be reinserted in any subsequent iteration.
The algorithm terminates when at least one of the following conditions is satisfied:
\begin{enumerate}
    \item $m' < k(m')$: in such a case the current solution takes just the first $m'$ rows (Line~\ref{code:2-last});
    \item $m' = 1$: which is the main condition of the algorithm.
\end{enumerate}

Even if more than $c$ positions of $\tilde{R}$ may be visited for each value of $m'$ (Line~\ref{code:2-general-add}), overall
during the second phase (Line~\ref{code:2-first-row}), 
no more than $|\tilde{R}|=m$ rows are visited. 
Since any row visited is handled in constant time, 
to update the solutions for all the values of $m'$ costs $\mathcal{O}(m)$.

The total cost of \ofri is then $\mathcal{O}(m\cdot \max\{n,\log m\})$ since
it spends $\mathcal{O}(mn + m \log m)$ for the pre-processing phase
and $\mathcal{O}(m)$ for the second phase.
Thus, the overall time complexity of \ofri is never worse than that of \ofr,
but ignoring the first pre-processing step whose results could be passed in input to the algorithm
along with the reward graph, 
\ofri is much faster than \ofr.
Indeed, \ofri costs $\mathcal{O}(m)$ time, while \ofr  $\mathcal{O}(m^2)$ time.

By the optimality of \ofr and by the above discussion the next theorem can be stated. 

\begin{theorem}
Algorithm \ofri
optimally solves \prob-FR in time $\mathcal{O}(m\cdot \max\{n,\log m\})$. 
\end{theorem}
\begin{proof}
By the optimality of \ofr and by the arguments provided in describing the algorithm, optimality is achieved.
\end{proof}

\subsection{Optimal Solution for \prob-SC}
For \prob-SC, 
since there is only one column in $A_C$,
the robot is forced to go back-and-forth on each selected row.
We devise a dynamic programming algorithm, called
\textsc{Opt Single-column} (\osc), that optimally solves \prob-SC in polynomial time.
Clearly, \osc sub-optimally solves \prob.

During the initialization,
we create two tables $T$ and $R$ of size $m \times (n+1)$ and $m \times (\frac{B}{2}+1)$, respectively.
Table $R$ has columns $j=0, 1, \ldots, \frac{B}{2}$.
We initialize table $T$ as follows:
$T[i,j] = \sum_{k=0}^{j} r(v_{i,k}) = \sum_{k=1}^{j} r(v_{i,k})$ for each $i,j$ which represents, 
fixed a row $i$, the cumulative reward up to the $j$-th column starting from the leftmost side, with $0 \le j \le n$.
This phase has cost $\mathcal{O}(mn)$.
In table $R$, let $R[i,b]$ be the largest reward that can be attained with  budget $2b$ 
visiting the first $i$ vertices of row $0$.
Let $S[i,b]$ be the last column visited in row $i$ to obtain the reward $R[i,b]$.

As for \prob-FR, \prob-SC there are some general properties that will be exploited by our dynamic programming solution:
\begin{description}
    \item[PA:] If $B\geq 2nm+2(m-1)$ then the robot can visit the whole graph.
    \item[PB:] An optimal solution must traverse the selected rows in increasing (or decreasing) order w.r.t. their indices. 
    \item[PC:] An optimal solution visits a row at most once. Two (or more) visit of the same row can be in fact replaced by the only visit that reaches the furthest vertex.
    \item[PD:] For each row the algorithm has to select the last vertex to visit. 
    If no vertex is visited, still $v_{i,0}$ has to be traversed
    to reach the subsequent row $i+1$.
\end{description}

The optimal solution that visits the first 
$j$ vertices in row $i$, with $j \ge 0$,  gains $T[i,j]$ reward from row $i$, 
and spends budget $2b-2j-2$ to reach the previous row $i-1$.
Namely, we reserve 2 units of budget to change row and $2j$ units to traverse row $i$.
We thus have the next recurrence.
The first row of table $R$ is defined as follows:
\begin{equation*}\label{eq:first-row}
\small{
	R[1,b] = 
		\begin{cases}
	       	T[1, b] & 0 \le b \le n-1\\
         	T[1, n] & n \le b \le \frac{B}{2}
	   	\end{cases}
}
\end{equation*}

Then, for each subsequent row $R[i,b]$, $i \ge 2$:
\begin{equation*}\label{eq:dp-recurrence}
\small{
	R[i,\!b]\!=\!\!
		\begin{cases}
            -\infty &  \!\!\!\!b\!<\!i\!-\!1 \\
            0 & \!\!\!\!b\!=\!i\!-\!1\\
          	\max\limits_{0 \le j \le \min\{b-i, n\}} \!\!\{\!R[i\!-\!1, b\!-\!j\!-\!1]\!+\!T[i, j]\!\} & \!\!\!\!b\!>\!i\!-\!1
    	\end{cases}
}
\end{equation*}
Note that $j \le \min\{b - i, n\}$ is obtained by observing that $j \le n$, $b - j - 1 \ge 0$ and more strictly
$b-j-1 \ge i-1$ to avoid $-\infty$ reward. 
If $b=i-1$ and $i \ge 2$, $R[i,b]$ is just $0$ because there is only enough budget to reach row $i$ traversing $c_0$.
Precisely, $R[i,b]= \max_{j=0} \{ R[i - 1, b - j - 1] + T[i, j] \}= R[i - 1,b - 1]$ and recursively back up to $R[1,0]=0$. 

Table $S$ is then filled recalling for each position the column that has given the maximum reward,
i.e., $S[i,b] \gets j = \argmax R[i,b]$.
Finally, the reward of the optimal solution with budget $\frac{B}{2}$ is found by calculating $\max_{1 \le i \le m} R[i, \frac{B}{2}]$
because
we do not know in advance the furthest row reached by the optimal solution.
The solution is computed in $O(m)$ tracing back the choices using the table $S$.
Note that with a single execution of  \osc,  we compute the optimal reward for any value $b$, with $0 \le b \le \frac{B}{2}$. That is, the maximum reward for budget $b$ is the maximum in column $b$ of table $R$.

The algorithm runs in time 
$\mathcal{O}(m  n  \frac{B}{2})$
plus $\mathcal{O}(m  \frac{B}{2})$ to retrieve the solution,
and takes $\mathcal{O}(m  n + m  \frac{B}{2})$ space.
Since by property PA budget $B$ is upper bounded by $\mathcal{O}(mn)$, then algorithm \osc is strictly polynomial in the size of the input.

By the above discussion and sub-optimality arguments, the next theorem can be stated.

\begin{theorem}
Algorithm \osc
optimally solves \prob-SC in time $\mathcal{O}(m  n  \frac{B}{2})$.
\end{theorem}

\begin{proof}
The running time is obvious. 
Let us define $ R(i,b) $ as the optimal profit to reach any vertex of row $i$ when the budget is $2 b$.
The solution $S(i,b)$ associated with $R(i,b)$ 
has maximum profit and must traverse the vertices $v_{1,0}, v_{2,0}, \ldots, v_{i,0}$ of $c_0$.

For the correctness we prove by induction.
For the base case, $i=1$ and any $b$, or any row $i$ and $b \le i\!-\!1$, 
the correctness follows from the above discussion.

Induction Step: When computing $ R[i,b] $ by induction hypothesis, we have that $R[i-1, b-j-1] $ for any $0 \le j \le \min\{ b-i, n-1 \}$ are already computed correctly.
 Since any optimal solution $S[i,b]$ visits the rows in increasing index order, traverses only once row $i$, up to any $c_j$ (recall that $j=0$ means that the row is not visited), and must traverse the vertices $v_{1,0}, v_{2,0}, \ldots, v_{i,0}$ of $c_0$, $S[i,b]$ is built starting from a sub-problem that traverses the vertices $v_{1,0}, v_{2,0}, \ldots, v_{i-1,0}$ and some vertices of row $i$. Then, $S[i,b]$ is based on a sub-problem that considers up to row $i-1$, leaves $1$ unit of budget to reach row $i$, and leaves $j$ units of budget to reach vertex $j$ in row $i$, for some $0 \le j \le b-i$.
 Hence, the value of $R[i,b]$ in Eq.~\eqref{eq:dp-recurrence} is correct.
 Induction Step: When computing $R[i,b]$ by induction hypothesis, we have that $R[i\!-\!1, b\!-\!j\!-\!1]$ for any $0 \le j \le \min\{ b\!-\!i, n\!-1 \}$ were already computed correctly.
 Note that any optimal solution $S[i,b]$ is built starting from a sub-problem that traverses the vertices $v_{1,0}, v_{2,0}, \ldots, v_{i-1,0}$ and some vertices of row $i$.
 In fact, such solution visits the rows in increasing index order, 
 traversing any row $i$ only once up to a $c_j$ (recall that $j=0$ means that the row is not visited), and must traverse the vertices $v_{1,0}, v_{2,0}, \ldots, v_{i,0}$ of $c_0$.
 Then, $S[i,b]$ is based on a sub-problem that considers up to row $i-1$ keeping aside enough budget to reach vertex $j$ in row $i$, i.e., $1 + j$ units of budget for some $0\!\le\!j\!\le\!b\!-\!i$.
 Hence, the value of $R[i,b]$ in Eq.~\eqref{eq:dp-recurrence} is correct.

 Now, assume by contradiction, that there exists a solution $S'[i,b] \not =  S[i,b] $  with cost $R'[i, b] > R[i, b]$,
 and that is the first time that Eq.~\eqref{eq:dp-recurrence} does not provide the optimum.
 Let vertex $v_{i,j}$ be the vertex reached by $S'[i,b]$ in row $i$. 
 The solution $S'[i,b]$ gains $T[i,j]$, and 
 $R'[i,b]-T[i,j]$ corresponds to the profit of a sub problem $S[i-1,b-j-1]$ for which $R[i-1, b-j-1]$ is optimal.
 That is, $R'[i,b]-T[i,j]=R[i-1, b-j-1]$ and thus $R[i,b]=R[i-1, b-j-1]+T[i,j]=R'[i,b]$.

 Finally, since we do not know in advance the furthest row that belongs to the optimal solution,
 the optimal solution with budget $B$ is found in column $b$ and precisely it is $\max_{1 \le i \le m} R[i, B]$.
 \end{proof}

\subsection{Heuristic solution for \prob}
We propose an heuristic, called \textsc{Heuristic General-case} (\hgc), that
combines the sub-optimal solutions of \prob returned by \ofri and \osc algorithms in three strategies:
$H1$, $H2$, and $H3$.
$H1$ prefers full rows, while $H2$ and $H3$ select
partial rows on both the left and right side 
plus some full rows to complete the tour:
$H3$ traverses only two full rows (i.e., the minimum number of rows to reach the right side and be back at $v_{1,0}$), 
while $H2$ replaces all sufficiently long partial rows with full rows.
Precisely:

\begin{description}
    \item[H1:]
    First \ofri is applied. If some budget $p$ there remains, then clearly $p<2(n+1)$. \osc is applied to select
    partial rows on both the sides of the aisle-structure $A$, constrained to $p$ and to the requirement to return in $v_{1,0}$. In particular,  the partial rows can be selected among
    all the rows
    of the left side not yet selected, and 
    among the rows of the right side
    that the robot passes in front traversing the tour created for \ofri. 
    
    It is worthy to note that the reward of $H1$ always dominates that of \ofri.

	\item[H2:] Here first \osc is applied with the full budget $B$, obtaining a temporary solution $S_L$ then again \osc is applied with the full budget $B$ but starting and returning in $v_{1,n}$, hence optimizing on the right side and obtaining another temporary solution $S_R$. Successively, $S_L$ and $S_R$ are analyzed so as to select an even number of full rows. These will be the rows where $S_L$ and $S_R$ combined together select at least $\frac n 2$ elements. If none of such rows exist, then bounds $\frac n 3$, $\frac n 4$, $\ldots$, will be considered just to select two rows. The final solution will then include the selected full rows. According to the residual budget, the solution will be completed with all possible partial rows similarly to $H1$;
	
	\item[H3:] Here firstly two rows are selected: the first one and the furthest row reachable with the budget $B$. Successively, \osc is applied with the residual budget selecting the partial rows as in $H1$.
 \end{description}

Heuristic \hgc then returns the best solution among $H1$, $H2$, $H3$, \ofri and \osc.

\section{Simulations}
\label{sec:simulations}

\begin{figure}[ht]
	\begin{subfigure}[b]{0.2\textwidth}
		\centering
		\includegraphics[scale=1]{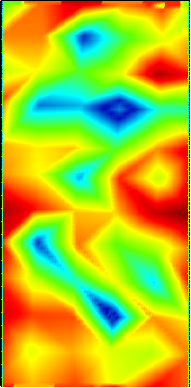}
		\caption{Real}
		\label{fig:heatmap_real}
	\end{subfigure}
	\centering
	\begin{subfigure}[b]{0.2\textwidth}
		\centering
		\includegraphics[scale=1]{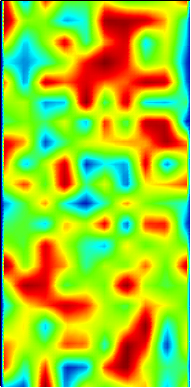}
		\caption{$\theta=0$}
		\label{fig:heatmap_theta_00}
	\end{subfigure}
	\begin{subfigure}[b]{0.2\textwidth}
		\centering
		\includegraphics[scale=1]{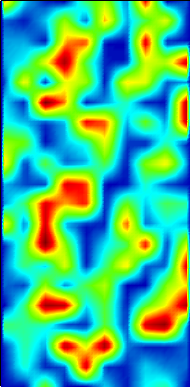}
		\caption{$\theta=0.8$}
		\label{fig:heatmap_theta_08}
	\end{subfigure}
	\begin{subfigure}[b]{0.2\textwidth}
		\centering
		\includegraphics[scale=1]{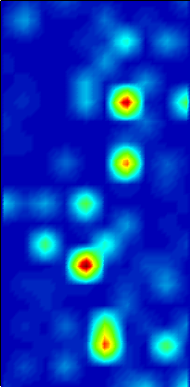}
		\caption{$\theta=1.9$}
		\label{fig:heatmap_theta_19}
	\end{subfigure}
	\begin{subfigure}[b]{0.2\textwidth}
		\centering
		\includegraphics[scale=1]{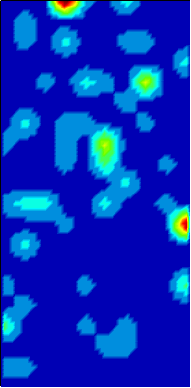}
		\caption{$\theta=2.7$}
		\label{fig:heatmap_theta_27}
	\end{subfigure}
	\caption{Reward maps for real and synthetic graph instances.}
	\label{fig:plot_compare_3}
\end{figure}

We test our algorithms
on the same benchmarks proposed in~\cite{thayer2018routing} that are generated from
a real world robotic precision irrigation application. 
The reward map of one of them is illustrated in
Fig.~\ref{fig:heatmap_real}, where the cold and hot colors represent low and high rewards, respectively.
We also test our algorithms
on synthetic instances
whose rewards are based on various Zipf distributions~\cite{tullo2003modelling}.
The synthetic instances consist 
of graphs $A(100,50)$ and $A(50,100)$.
The rewards are random integers in the interval $[0,100)$
that follow a Zipf distribution of parameter $\theta=\{0,0.9,1.8,2.7\}$.
To avoid a punctiform reward in $A$ and make it more continuous as it is in the real world,
we generate $\frac{m \cdot n}{5 \cdot 5}=\frac{50 \cdot 100}{5 \cdot 5}$ random numbers and we assign each of them to a $5 \times 5$ sub-graph.
When $\theta=0$, the rewards are uniformly distributed in $[0,100)$ (see Fig.~\ref{fig:heatmap_theta_00}),
while when $\theta$ increases the lowest rewards become more and more frequent
(see Figs.~\ref{fig:heatmap_theta_08}-\ref{fig:heatmap_theta_27}).

\begin{figure*}[t]
	\centering
	\begin{subfigure}[b]{0.3\textwidth}
		\centering
		\includegraphics[scale=1]{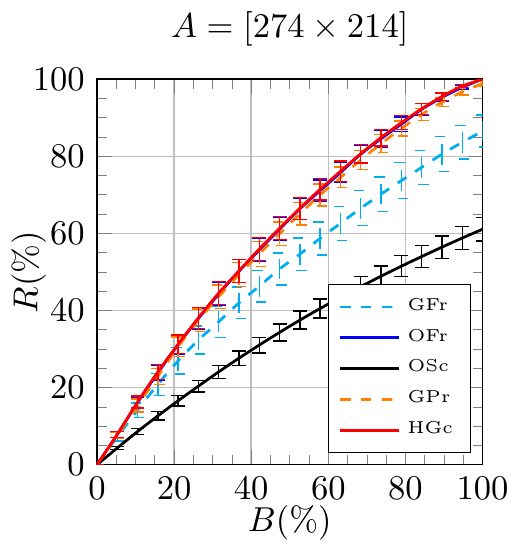}
		\caption{Real}
		\label{fig:plot_compare_real}
	\end{subfigure}
	\begin{subfigure}[b]{0.3\textwidth}
		\centering
		\includegraphics[scale=1]{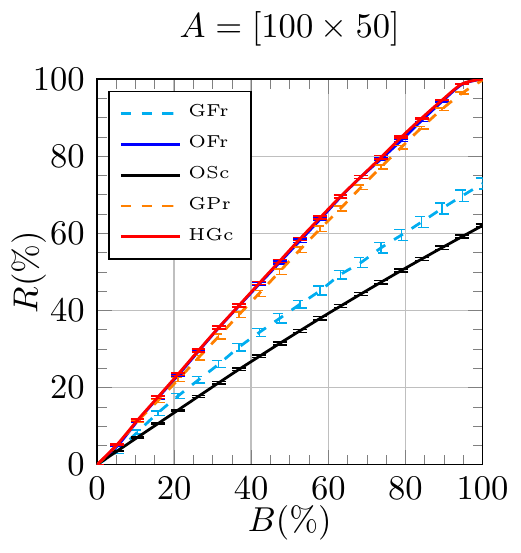}
		\caption{$\theta=0$}
		\label{fig:plot_compare_100}
	\end{subfigure}
	\begin{subfigure}[b]{0.3\textwidth}
 		\centering
		\includegraphics[scale=1]{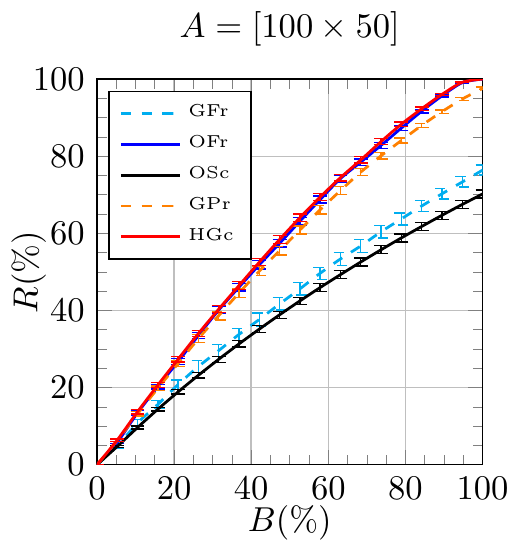}
 		\caption{$\theta=0.8$}
 		\label{fig:plot_compare_300}
 	\end{subfigure}
	\begin{subfigure}[b]{0.3\textwidth}
		\centering
		\includegraphics[scale=1]{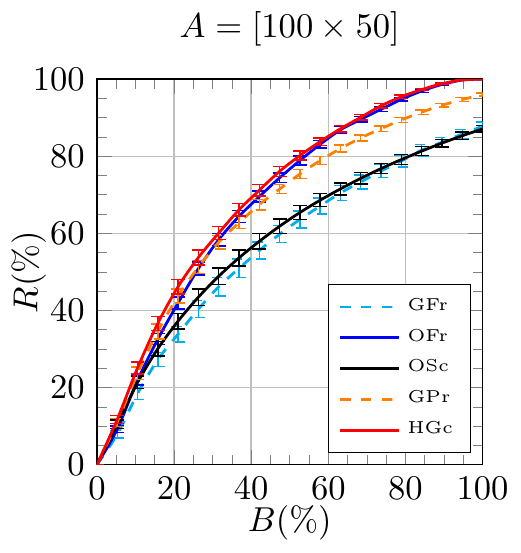}
		\caption{$\theta=1.9$}
		\label{fig:plot_compare_500}
	\end{subfigure}
	\begin{subfigure}[b]{0.3\textwidth}
		\centering
		\includegraphics[scale=1]{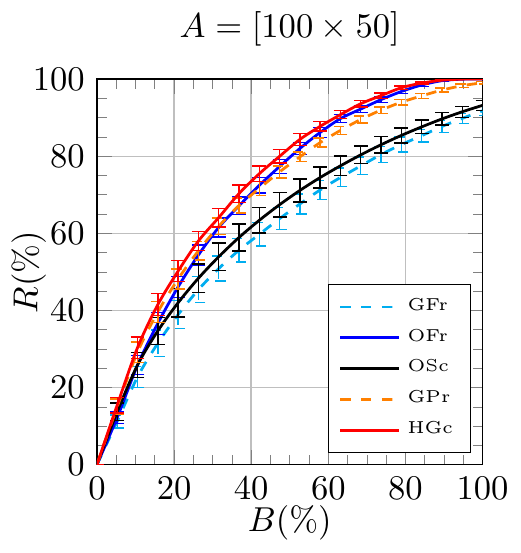}
		\caption{$\theta=2.7$}
		\label{fig:plot_compare_700}
	\end{subfigure}
	\caption{The performances for the real benchmark and synthetic data with different values of $\theta$. 
	}
	\label{fig:plot_compare_all}
\end{figure*}

For each $A$ and $\theta$, we run our new algorithms \ofr, \osc, \hgc and contrast them with 
the two greedy heuristics \textsc{Greedy Full-Row} (\gfr) and \textsc{Greedy Partial-Row} (\gpr) presented in~\cite{thayer2018routing}. Both \ofr and \ofri optimally solve \prob-FR, so we do not need to test both.

The \gfr and \gpr heuristics choose a subset of full and partial rows to be traversed, respectively.
At each selection, the robot computes the budget required to collect
rewards from its current position, and prefers row/vertices with maximum reward per unit of budget.
The time complexities of \gfr and \gpr are $\mathcal{O}(m^2)$ and $\mathcal{O}(m^2n)$, respectively.
Fixed $m$ and $n$ (i.e., the size of $A$) and $\theta$, we populate $30$ random graphs.
For each graph, we test each algorithm with budget $B$ increasing in $20$ steps from the minimum to the maximum value (see \textit{P1}, Sec.~\ref{sec:full-rows-policy}).
Since the overall reward differs for each graph (as rewards are randomly generated),
we return the reward gained as a percentage of the overall reward.
Finally,
we plot the average of the results on the $30$ instances, along with their $95\%$ confidence interval.
Fig.~\ref{fig:plot_compare_all} compares the reward over the different algorithms
on the real and synthetic data and different  $\theta$.  
When $\theta=0$, the performance of \ofr increases linearly with the budget
and it is absolutely comparable with that of the general heuristics \gpr and \hgc  even if it is computationally  lighter. 
\gfr instead performs poorly with respect to \ofr.
\osc is the worse in the uniform case and not surprisingly it gains about half of the reward of \ofr. 
In fact, \osc requires the double of the budget than \ofr to explore a single row. 
When $\theta$ increases, and rewards become scarcer,
the percentage of gained reward increases faster than the budget: at $\theta=1.9$ with $50\%$ of the budget,
\ofr gains already more than $70\%$ of the reward.
The performances of \ofr is coincident with that of \hgc up to $\theta=1.9$,
and about $1$--$2\%$ less when $\theta=2.7$.
For large values of $\theta$, the performance of \gpr slightly downgrades especially for budget between
$50\%$ and $80\%$, whereas that of \osc upgrades although \osc explores only the left side of $A$.
In fact,  when the reward is unbalanced, for \osc  it is easier to decide
which vertices to take,
while \gpr may waste budget on useless vertical movements because the rewards are widely
scattered.
\osc reaches the same performance of \hgc and \gpr when $\theta \ge 1.9$  and $B \le 20\%$.

Fig.~\ref{fig:plot_compare_real} compares the algorithms behavior on a set of $20$ real 
world robotic precision irrigation graphs, collected in the open fields by the authors of~\cite{thayer2018arouting}.
\hgc and \ofr achieve the best results,
and generally improves by less than $2\%$ the performance 
of \gpr (with a peak of $5\%$ when $B< 20\%)$.
In general, the \gfr heuristic performs better on the real data than on the synthetic ones.
Vice-versa, \osc performs better on the synthetic data than on the real ones.
According to our experiments, the performance on real graphs are comparable with that of synthetic graphs
of moderate skewness. 

\subsubsection*{Observations}
Although all experiments show excellent performance for \gpr and even better for \ofr, we are aware that one can hand-pick graphs where such algorithms behaves rather poorly compared the optimal solution. 
For example, consider a graph 
$A = A(m,n)$ built as follows: 
for each vertex $v_{i,2}$ with $i \in 1, \ldots, m-1$ 
we set $r(v_{i,2}) = 2i - \epsilon$; 
we set reward $r(v_{m,m-2}) = m-1 + 2(m-2) = 2m - 3$ for the vertex $r(v_{m,m-2})$ of the last row; 
and finally we set reward equal to zero for all the other vertices.
Let $B=2(m-2) + 2(m-1)$ the given budget.
The \ofr will select the furthest row $m'$  it can reach
constrained on $B$ and the previous row $m'-1$ for a total reward that 
cannot be larger than $2m-5+m-\epsilon< 3m -\epsilon$.
The \gpr will choose in the first iteration the vertex $v_{m,m-2}$ consuming the whole budget 
for a total reward of $r(v_{m,m-2}) = m-1 + 2(m-2) = 2m - 5$.
The \osc algorithm will compute a solution composed by the first $m-1$ rows with a total reward 
of $\sum_{i=2}^{m - 2}(i - \epsilon) = \frac{1}{2} (m - 3) (m - 2 \epsilon)$.
Since the optimal solution will gain at least the same reward as \osc, both \ofr and \gpr are away from the optimum by 
an arbitrarily large factor $m$.

Similar examples can be found where \osc looses, while \gpr and \ofr are much closer to the optimum.
In conclusion, since \osc and \ofr seem to show opposite behaviors in the same instance (like in the example above)
and since \hgc applies both,
our aim for \hgc is to limit  performance losses. On the one hand we could not find any dramatic instance for \hgc as for the other heuristics. On the other hand we are looking forward for formal arguments to guarantee some limited approximation ratio. 

\section{Conclusion}
\label{sec:conclusions}
We have shown how to optimally solve in polynomial time 
the two special cases \prob-FR and \prob-SC of 
\prob.
We have also simulated the new optimal proposed algorithms \ofr  and \osc as well as the new heuristic \hgc in the special vineyard context
where \prob was originally defined.
Although \hgc gains in percentage few units over the previous best algorithm,  it can be still valuable
in high constrained scenario.
As future work, we will undertake further investigations for providing formal guarantees on the quality of the solutions in the general case.

\bibliographystyle{IEEEtran}
\bibliography{IEEEabrv,bib-icra-20,bib-carpin}

\end{document}